\documentclass[aps,prl,twocolumn,groupedaddress,showpacs]{revtex4}
\pdfoutput=1
\usepackage{graphicx}
\usepackage{amsmath}

\begin{document}

\title{Measurement of Growing Dynamical Lengthscales and Prediction of the Jamming Transition in a Granular Material}

\author{Aaron S. Keys$^{ 1 \dag } $}
\author{Adam R. Abate$^{ 2 \dag }$}
\author{Sharon C. Glotzer$^{1,3}$}
\author{Douglas J. Durian$^2$}
\affiliation{$^1$Department of Chemical Engineering, University of Michigan, Ann Arbor, Michigan 48109-2136 \\
$^2$Department of Physics and Astronomy, University of Pennsylvania, Philadelphia, Pennsylvania 19104 \\
$^3$Department of Materials Science Engineering, University of Michigan, Ann Arbor, Michigan 48109-2136 \\
$^\dag$ These authors contributed equally to this work.}

\date{\today}

\begin{abstract}
Supercooled liquids and dense colloids exhibit anomalous behaviour known as ``spatially heterogeneous dynamics'' (SHD), which becomes increasingly pronounced with approach to the glass transition~\cite{ediger00, kegel00, weeks2000}.  Recently, SHD has been observed in confined granular packings under slow shear near the onset of jamming, bolstering speculation that the two transitions are related~\cite{pouliquen2003fluctuating, marty05, dauchot05}.   Here, we report measurements of SHD in a system of air-driven granular beads, as a function of both density and effective temperature.  On approach to jamming, the dynamics become progressively slower and more spatially heterogeneous.  The rapid growth of dynamical time and length scales characterizing the heterogeneities can be described both by mode-coupling theory~\cite{gotze92} and the Vogel-Tammann-Fulcher (VTF) equation~\cite{vogel21}, in analogy with glass-forming liquids. The value of the control variable at the VTF transition coincides with point-J~\cite{o2002random, o2003jamming} the random close-packed jamming density at which all motion ceases, indicating analogy with a zero temperature ideal glass transition.  Our findings demonstrate further universality of the jamming concept and provide a significant step forward in the quest for a unified theory of ``jamming'' in disparate systems.
\end{abstract}

\maketitle

At low temperature, high density, and low driving, the constituent particles in supercooled liquids~\cite{ediger00}, dense colloids~\cite{kegel00, weeks2000}, and granular packings~\cite{pouliquen2003fluctuating, marty05, dauchot05}, respectively, are nearly locked into a single disordered configuration in which motion is heterogeneous in space and time. Dynamics in these systems may be governed by proximity to a generic ``jamming transition~\cite{liu98}'', beyond which rearrangements cease and the viscosity diverges.  Key features of SHD on approach to the transition include unusual correlations~\cite{bennemann99} in which particles move in one-dimensional paths (``strings~\cite{donati98}'') that aggregate into clusters~\cite{donati99}, and dynamical correlations as measured by a dynamic four-point susceptibility $\chi_4$~\cite{donati02, glotzer00x4, chandler2006}.  Clusters of strings arise naturally in dynamic facilitation~\cite{cgprl03, cgpnas03} theory and the random first-order theory of glasses~\cite{stevenson2005}; their shape reflects the fractal nature of dynamical motion in these systems~\cite{stevenson2006shapes}.  Strings are also a crucial ingredient in a recent theory of liquid dynamics near the glass transition~\cite{langer06}. 

Recent studies demonstrate that close-packed granular systems under slow shear exhibit SHD as well~\cite{pouliquen2003fluctuating, marty05, dauchot05}, bolstering speculation that liquids and granular matter share dynamical similarities on approach to the jammed state. However, the universality of the jamming hypothesis has not yet been tested in terms of variation in the hallmark dynamical heterogeneities as a function of the control parameter.  Here, we present the first simultaneous measurements in any experimental system of the growth of the cluster correlation length, string length, four-point correlation length, and their characteristic timescales by varying the control parameter. We show that the SHD observed in a far-from-equilibrium, athermal system of air-fluidized granular beads is essentially indistinguishable from that observed in thermal systems like supercooled liquids and dense colloidal suspensions. Moreover, we show that theoretical models developed for the glass transition can be used to describe our granular system, and predict a mode-coupling like transition and, more importantly, the jamming transition packing fraction, known as point-J~\cite{o2002random, o2003jamming} from quantities characterizing SHD. 

\begin{figure*}
\begin{center}
\includegraphics[width=0.95\textwidth]{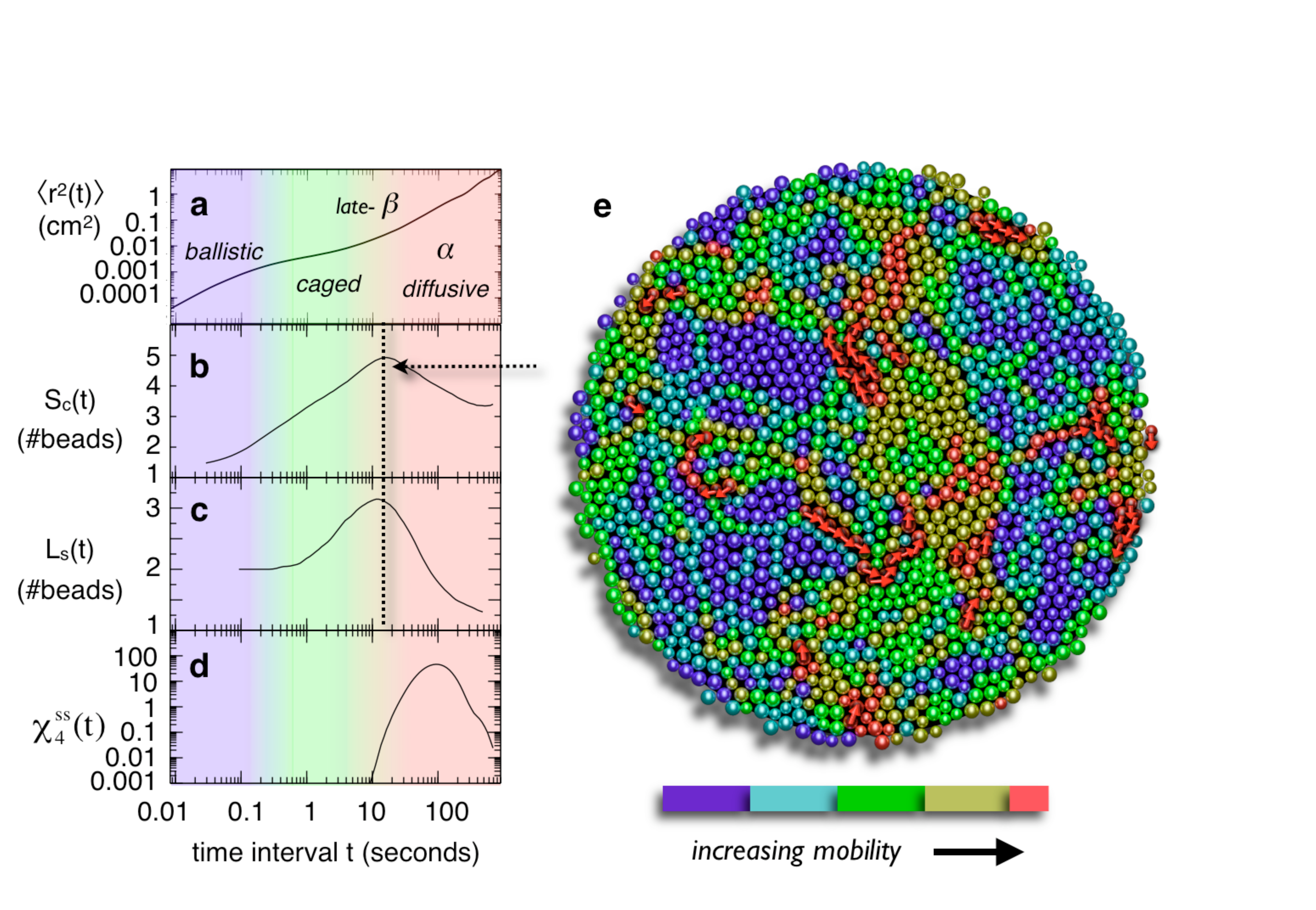}
\caption[Bead Dynamics at Area Fraction $\phi=0.773$ as a Function of Delay Time]{\label{fig:Dynamics773} (a) Mean square displacement. (b) Number average mobile bead cluster size $S_c(t)$. c) Number average string length $L_s(t)$. (d) Self contribution to the four point susceptibility $\chi^4_{ss(t)}$.  (e) An instantaneous bead configuration where the colouring of beads indicates the mobility over a time interval of 12s (the timescale for both maximum cluster size and string length).  The 10\% most mobile beads are red; note that they form clusters. Beads moving in strings have vectors superimposed to indicate their directional motion.  Note that the dynamics are spatially heterogeneous.}
\end{center}
\end{figure*}

We characterize the spatiotemporally heterogeneous nature of dynamics in an athermal, far-from-equilibrium system of air-driven steel spheres on approach to jamming.  Compared to sheared or shaken granular systems, in which energy is injected at the boundaries, in air-driven systems the energy input is uniform in space and time.  Our granular system consists of a 1:1 bidisperse mixture of steel beads of diameters $d_s = 0.318$ cm and $d_l = 0.397$ cm, with respective masses of 0.130 gm and 0.266 gm, confined to a circular region of diameter 17.7 cm.  The packing density is varied from an area fraction of $\phi = 0.597$ to $\phi = 0.773$ by changing the total number of beads from 1470 to 1904.  Bead motion is restricted to rolling within a horizontal plane, and is excited by an upward flow of air at a fixed superficial flow speed of 545 cm/s. Bead positions are identified by reflecting light from their chrome surface to a camera three feet above.  The duration of experimental runs is 20 minutes.  By contrast with the molecules in a supercooled liquid, here the particles are macroscopic objects driven at random by a continuous input of energy.  Consequently the speed distributions are non-Maxwellian, and the average kinetic energies of the two bead species are unequal.  Nevertheless, as reported previously~\cite{abate2006approach}, the system mimics a simple liquid for low $\phi$ and exhibits tell-tale changes in the average structure and dynamics at increasing packing densities.

\begin{figure*}
  \begin{center}
    \begin{minipage}[m]{0.6\linewidth}
      \includegraphics[width=\linewidth]{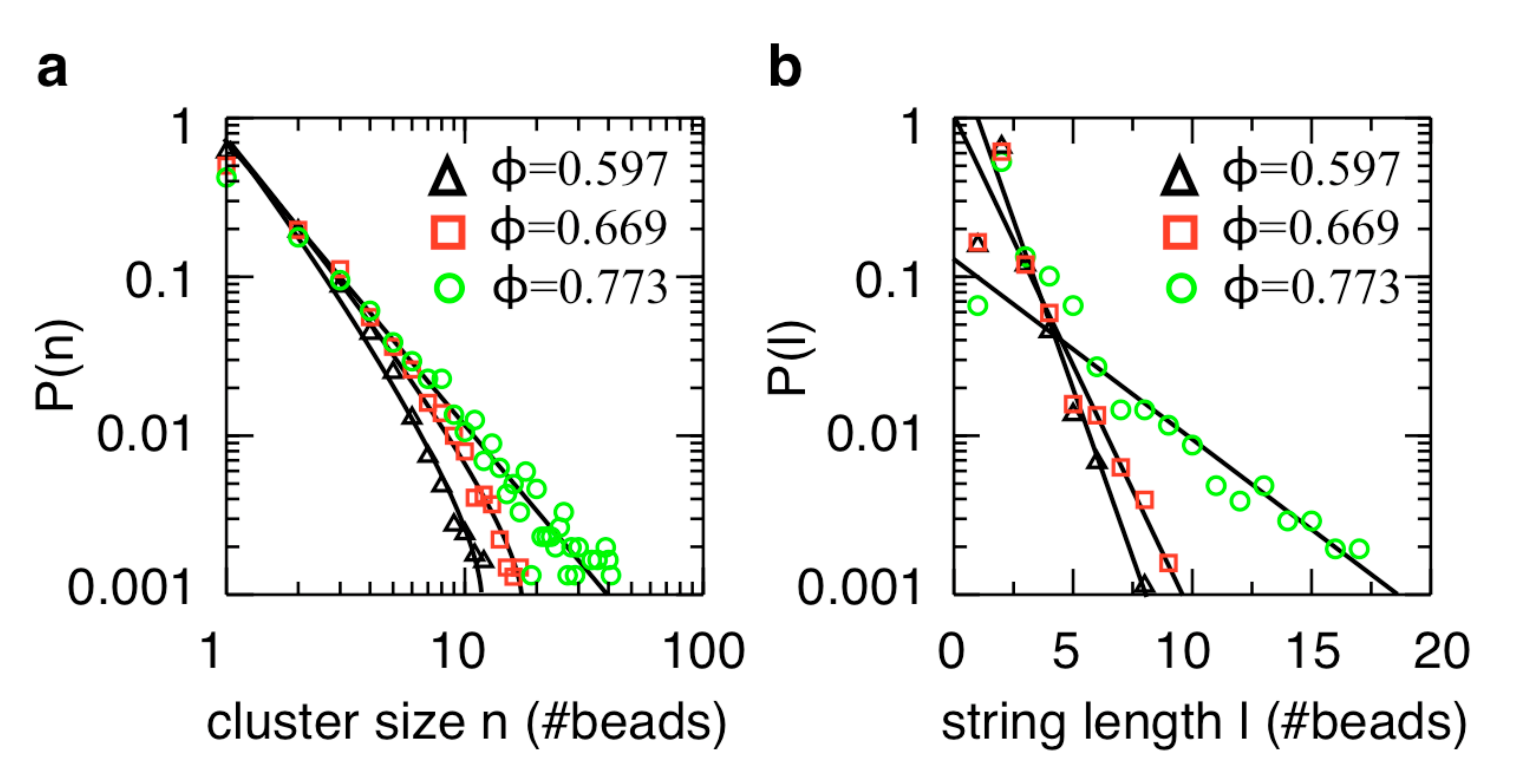}
    \end{minipage}\hfill
    \begin{minipage}[m]{0.4\linewidth}
\caption[Size Distributions of Clusters and Strings]{\label{fig:SizeDistributions} (a), (b) Distribution of (a) cluster sizes and (b) string lengths for three values of control variable $\phi$.  The cluster size distribution approaches a power law with increasing $\phi$ (solid lines indicate fit by power law multiplied by an exponential cutoff), while the string length distribution is exponential.}
\label{fig:SizeDistributions}
    \end{minipage}
  \end{center}
\end{figure*}

Dynamical characteristics for an example case, $\phi = 0.773$, are displayed in Fig.~\ref{fig:Dynamics773}a-d.  The mean-squared displacement versus time interval (delay time), Fig.~\ref{fig:Dynamics773}a, averaged over all start times and all beads, is perhaps the most familiar quantity.  It shows ballistic motion at short times, diffusive motion at late times, and a plateau of sub-diffusive motion at intermediate times that is indicative of caging.  While informative, the mean-squared displacement cannot distinguish uniform from heterogeneous dynamics.  For this, we perform three measurements developed for supercooled liquids.  The first involves clusters of beads that are ``mobile,'' i.e. which have displacements ranking among the top 10\% of all bead displacements in a given delay time~\cite{kob1997} (in this system, 10\% gives the largest distinction between mobile and immobile beads at all packing densities).  The configuration displayed in Fig.~\ref{fig:Dynamics773}e, where beads are colour-coded according to mobility, demonstrates that the mobile beads are not distributed at random; rather, mobile and immobile beads are clustered and spatially separated, indicating spatially heterogeneous dynamics.  One measure of SHD is thus the average size of mobile clusters, $S_c(t)$, defined as the average number of neighbouring mobile beads for a given time interval $t$.  The motion within a mobile cluster, shown by the displacement vectors in Fig.~\ref{fig:Dynamics773}e, tends to be correlated into quasi-1d paths called ``strings~\cite{donati98}.'' Thus a second measure of SHD is the average string length, $L_s(t)$, defined in terms of the number of beads that, within a time interval t, replace the initial positions of neighbouring beads to within a tolerance of $0.3 d_s$~\cite{donati98}.  Yet a third measure of SHD may be constructed from a four-point susceptibility $\chi_4(t)$, which measures the extent to which the dynamics at any two points in space are correlated within a time interval~\cite{donati02}.  The self-contribution $\chi_4^{SS}(t)$ dominates the general result~\cite{glotzer00x4} and is computed from the variance of the self overlap order parameter $q_S(t)$, which decays from one to zero:  
\begin{equation}
\chi_4^{SS}(t) = N \left[ \left< q_S(t)^2 \right> - \left< q_S(t)^2 \right> \right].
\end{equation}
Here, $N$ is the total number of beads, and $q_S(t)$ is defined as:
\begin{equation}
q_S(t) = \frac{1}{N} \sum_{i=1}^N w\left( \left| \textbf{r}_i(t) - \textbf{r}_i(0) \right| \right). 
\end{equation}
The overlap parameter $w$ is defined by:
\begin{equation}
w = \begin{cases}
1 & \left| \textbf{r}_i(t) - \textbf{r}_i(0) \right| < 0.5d_s\\
0 & \left| \textbf{r}_i(t) - \textbf{r}_i(0) \right| \geq 0.5d_s.
\end{cases}
\end{equation}
Here, $\textbf{r}_i(t)$ is the position of bead $i$ at time $t$; averages are taken over all beads and over all start times.

The example results in Fig.~\ref{fig:Dynamics773}b-d for the cluster size $S_c(t)$, the string length $L_s(t)$, and the four-point susceptibility $\chi^4_{ss}(t)$, all exhibit well-defined peaks as a function of time interval, as found in glass-forming liquids.  The locations of the peaks indicate the time interval over which the dynamics are most heterogeneous, and the heights of the peaks indicate the spatial extent or ``strength'' of the heterogeneities.  As with glass-forming liquids~\cite{donati99, gebremichael2001spatially, yeshi04, vogel04} and colloids~\cite{weeks2000}, the cluster size and string length are largest at the crossover between caged and diffusive motion, while $\chi^4_{ss}(t)$ (and $\chi^4 (t)$) peaks later, in the so-called alpha or structural relaxation regime~\cite{glotzer00x4, naida}.  The athermal air-fluidized beads therefore exhibit spatially heterogeneous dynamics that is identical to thermal glass-forming systems with respect to these three measures.   

Now that spatially heterogeneous dynamics are established for gas-fluidized beads, we turn to their variation as a function of control parameter. The distribution of cluster sizes at the peak time interval, shown in Fig.~\ref{fig:SizeDistributions}a for three different packing densities $\phi$, approaches a power-law   as $\phi$ is increased. This is consistent with the percolation of mobile bead clusters; similar power-laws have been observed in colloids~\cite{weeks2000} and in simulations of supercooled liquids~\cite{donati99, gebremichael2001spatially, yeshi04}  near the mode-coupling temperature.  Furthermore, the distribution of string lengths at the peak time interval, shown in Fig.~\ref{fig:SizeDistributions}b, is exponential,  , at all values of $\phi$ , where $l_0$ is set by the average string length.  This is consistent with behaviour reported in simulations of several supercooled liquids~\cite{donati98, yeshi04, aichele2003polymer}.  We note the average cluster size is not much larger than the average string length, although the largest clusters observed (~100 particles) are substantially larger than the largest string observed (~30 particles) (not shown).

\begin{figure}
\begin{center}
\includegraphics[width=0.9\columnwidth]{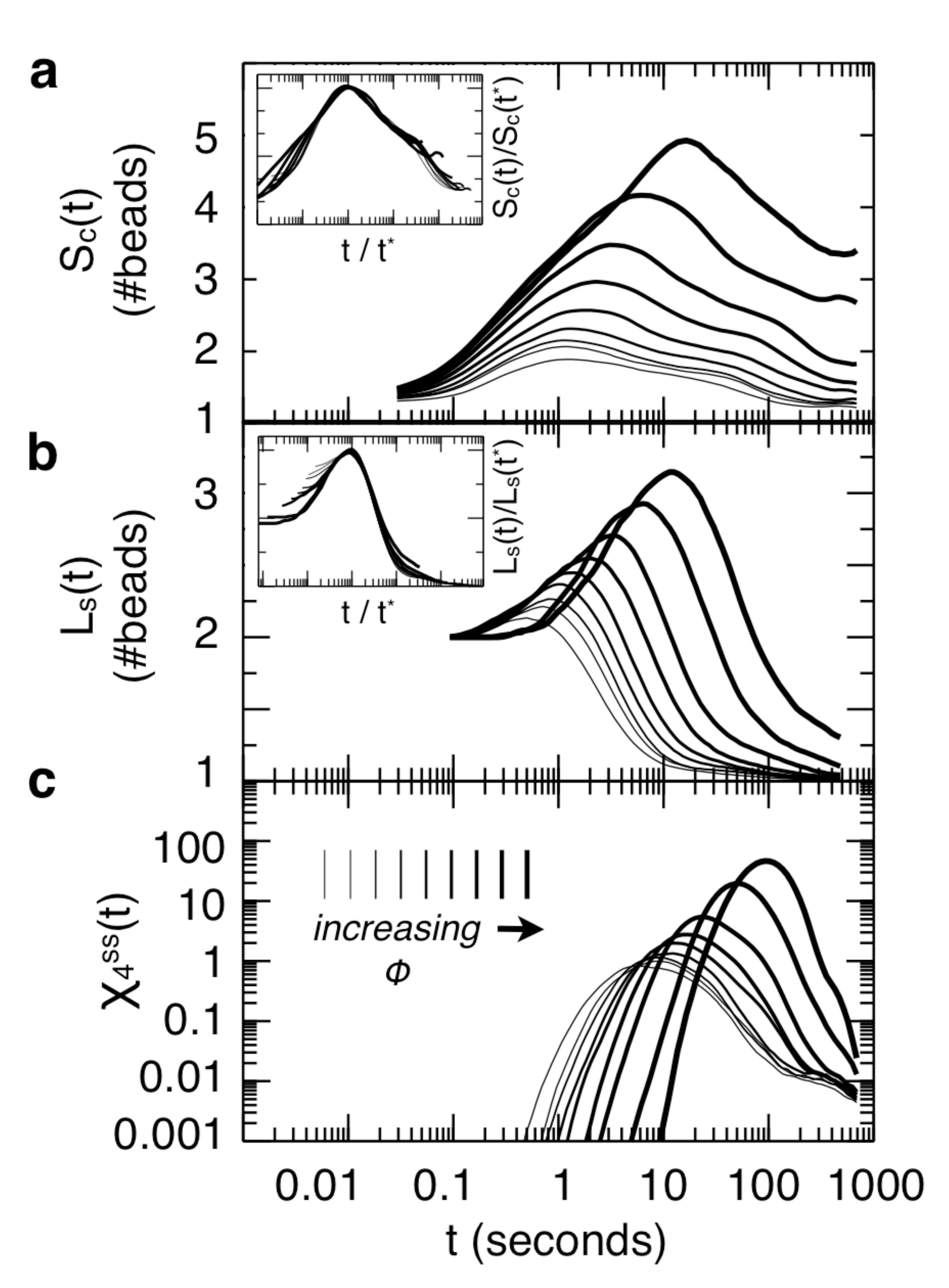}
\caption[Variation of Spatially Heterogeneous Dynamics (SHD) as a Function of Control Variable]{\label{fig:SHDControl} (a) Cluster size, (b) string length, and (c) self contribution to the four-point susceptibility, all as a function of time interval t, for a sequence of area fractions $\phi$.  In order of increasing peak height: $\phi=0.597, 0.633, 0.647, 0.669, 0.693, 0.722, 0.742, 0.762, 0.773$.  Insets in (a) and (b) show collapse of all data sets upon scaling by the peak heights and peak times.}
\end{center}
\end{figure}

Results for $S_c(t)$, $L_s(t)$, and $\chi^4_ss(t)$ vs $t$ are displayed in Fig.\ref{fig:SHDControl}a-c for a sequence of different packing densities $\phi$.  When beads are added to the system, the average effective temperature also decreases, resulting in a trajectory in the ($\phi$ ,$T_{\mathit{eff}}$) phase diagram that heads towards point-J, the zero-temperature jamming transition previously found for this system at $\phi =0.83$, which is coincident with the packing density at which the system is random close-packed.  As the motion becomes more restricted, the peaks in all three measures of SHD grow and move to later times.  Therefore, the dynamics not only slow down but also become more heterogeneous on approach to point-J.  Since the SHD functions have approximately the same shape when viewed on a log-log plot (see data collapse in insets of Fig.~\ref{fig:SHDControl}a,b), this behaviour is fully characterized by the $\phi$-dependence of the characteristic or peak time scales  $\{ t^*_{Sc}, t^*_{Ls}$ and $t^*_{\chi 4} \}$ and length scales $\{ \zeta_{Sc}(t^*_{Sc}), \zeta_{Ls}(t^*_{Ls})$ and $\zeta_{\chi 4} (t^*_{\chi4})\}$.  The length $\zeta_{Ls}(t^*_{Ls}) \propto L_s$ is a correlation length for stringlike motion, $\zeta_{Sc}(t^*_{Sc}) \propto S_c$ is a correlation length of mobile particle clusters, and  is a correlation length~\cite{dauchot05} of clusters of caged particles.  On approach to point-J, both the characteristic times and the correlation lengths appear from Fig.~\ref{fig:SHDControl} to grow without bound.  This is reminiscent of behavior for supercooled liquids as temperature is lowered.  Though very different, both types of systems appear to approach an unusual critical point where the growing length scale is purely dynamical, such that there is no macroscopic change in instantaneous structure~\cite{ediger00, stone2004stress, corwin2005structural, silbert2005vibrations}.

\begin{figure}
\begin{center}
\includegraphics[width=\columnwidth]{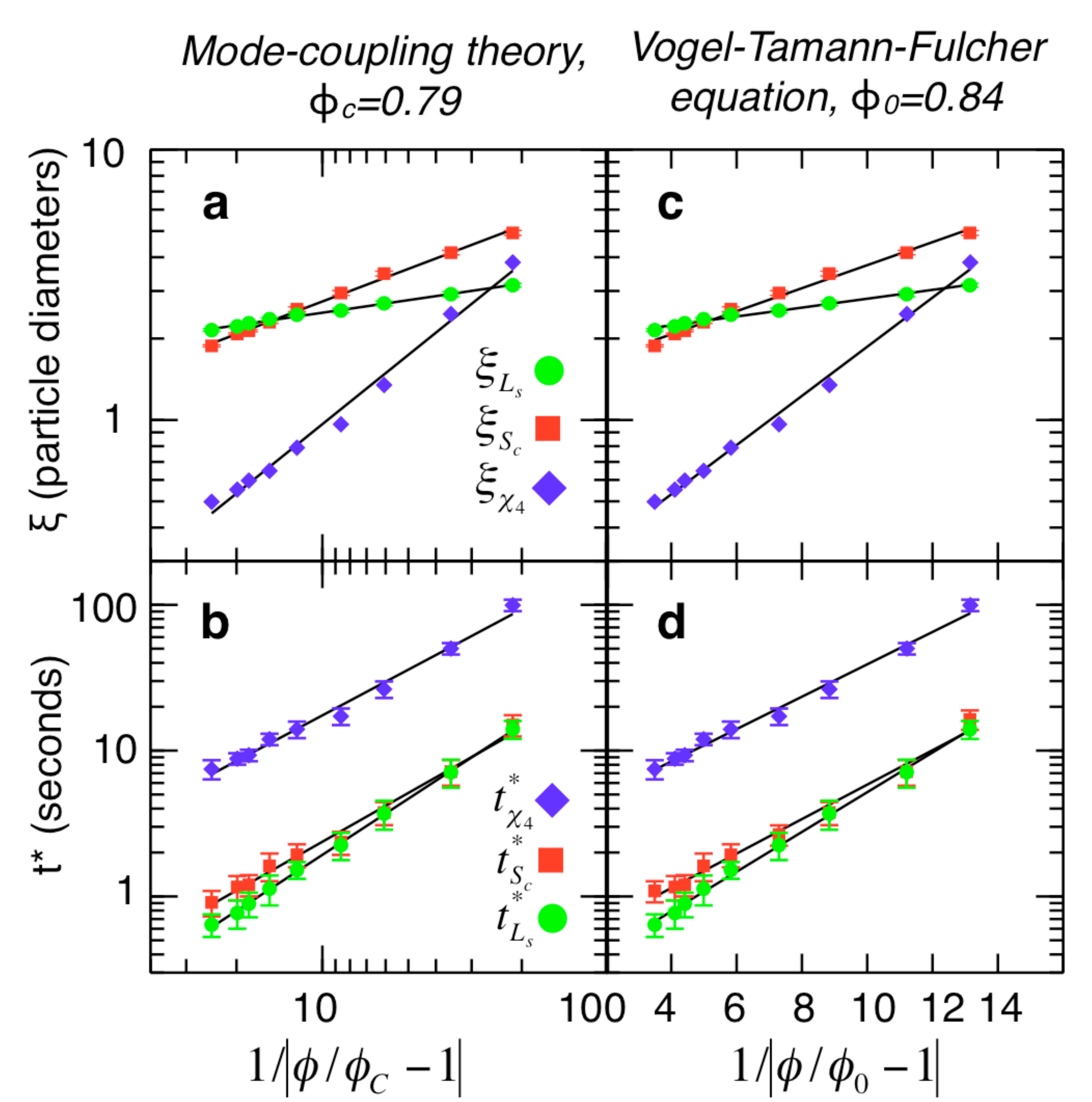}
\caption[Dependence of Dynamic Time Scales and Length Scales on Packing Density $\phi$]{\label{fig:MCTVTF} (a) Log-log plot of  dynamical correlation lengths versus $1/|\phi/\phi_c-1|$ fit with $x=0.40$ for $\zeta_{Sc}$, $x=0.15$ for $\zeta_{Ls}$ and x=1.70 for $\zeta_{\chi 4}$.  (b) Log-log plot of characteristic times $t^*$ versus $1/|\phi/\phi_c-1|$ with $x=1.3$ for $t^*_{Sc}$, $x=1.1$ for $t^*_{Ls}$, and $x=1.0$ for $t^*_{\chi 4}$. (c) Semilog plot of dynamical correlation lengths versus $1/|\phi/\phi_0-1|$ fit with $E=0.10$ for $\zeta_{Sc}$, $E=0.03$ for $\zeta_{Ls}$, and $E=0.42$ for $\zeta_{\chi 4}$.  (d) Semilog plot of all $t^*$‚Äôs vs $1/|\phi/\phi_0-1|$ fit with E=0.27.  See text for fit expressions. Error bars for (a), (c) represent the standard error of the measurement.  Error bars for (b), (d) represent the uncertainty in determining the time at which the function is maximum. .}
\end{center}
\end{figure}

To further quantify this analogy, the growth of the characteristic timescales and dynamical length scales is shown in Fig.~\ref{fig:MCTVTF}a-d as a function of packing density.  Motivated by recent studies~\cite{biroli2004, birolimct1, birolimct2} predicting a power law divergence of dynamical lengthscales from mode coupling theory~\cite{gotze92} (MCT), as well as earlier applications of MCT to liquids and colloids, we fit all data to a power-law of the form $1/ \left| \phi/\phi_c \right|^x$, where both $\phi_c$ and $x$ are adjustable parameters.  As seen in Fig.~\ref{fig:MCTVTF}a,b, excellent fits are obtained to all data for a single value $\phi_C = 0.79 \pm 0.02$.  This value of $\phi$ lies well above the onset of caging and is less than the jamming packing fraction, in analogy with well-established findings that the mode-coupling temperature is below the caging transition but above the glass transition temperature~\cite{gotze92, reichmanmct}, and demonstrates for the first time a mode-coupling-like transition in a granular system.  In addition to MCT, the glass transition can also be well-described by the Vogel-Tammann-Fulcher (VTF) equation~\cite{vogel21}; therefore, we also fit the characteristic time and length scales to the form $\exp \left( E/ \left| \phi/\phi_0 - 1 \right| \right) $ , where $E$ and $\phi_0$ are adjustable parameters.  As seen in Fig.~\ref{fig:MCTVTF}c,d, excellent fits are obtained to all data for a single value $\phi_0 = 0.84 \pm 0.02$, consistent with the value of random-close packing for the bead system and the value of point-J.  Since random close packing is the densest random packing possible and the point at which all motion ceases, a VTF packing fraction of $\phi_{RCP}$ is analogous to an effectively zero-temperature ideal glass transition, consistent with the definition of point-J.  This is the first prediction of point-J in a granular system from analysis of spatially heterogeneous dynamics.

Our study implies that the behaviour of jammed systems, both thermal and athermal alike, may be understood using the theoretical tools developed for liquids.  This, in turn, highlights the importance of packing in the underlying physics of the glass transition and jamming. Our results open the door to future theoretical insight into the relationship between granular materials and supercooled liquids, which might be united by a unified theory of jamming.

\textbf{\textit{Acknowledgements:}} This work was supported by the National Science Foundation under grant no. NSF-DMR0514705 (A.R.A. and D.J.D.), NASA under grant no. NNC04GA43G (A.S.K. and S.C.G.) and the Department of Education GAANN fellowship program (A.S.K.). 
\bibliography{shd}

\end{document}